\documentclass[12pt,preprint]{aastex}

\def\ergs{ergs s$^{\rm -1}$ }
\def\flux{ergs s$^{\rm -1}$ cm s$^{\rm -2}$ }

\def\msolar{\mbox{M$_{\odot}$} }

\def\ltsima{$\; \buildrel < \over \sim \;$}
\def\simlt{\lower.5ex\hbox{\ltsima}}            % < over ~
\def\gtsima{$\; \buildrel > \over \sim \;$}
\def\simgt{\lower.5ex\hbox{\gtsima}}            % > over ~

\def\beq{\begin{equation}}
\def\eeq{\end{equation}}
\def\beqary{\begin{eqnarray}}
\def\eeqary{\end{eqnarray}}
\newcommand{\eqref}[1]{(\ref{eq:#1})}

\begin{document}


\title{ Discovery of Diffuse X-ray Emission in One of
        the Nearest Massive Star-Forming Regions NGC 2024}


\author{Y. Ezoe\altaffilmark{1,2},
        M. Kokubun\altaffilmark{1},
        K. Makishima\altaffilmark{1,3},
        Y. Sekimoto\altaffilmark{4},
        K. Matsuzaki\altaffilmark{2}}


\altaffiltext{1}{Department of Physics, University of Tokyo,
                 7-3-1 Hongo, Bunkyo-ku, Tokyo, Japan}
\altaffiltext{2}{The Institute of Space and Astronautical Science,
                 3-1-1 Yoshinodai, Sagamihara, Kanagawa 229-8510, Japan}
\altaffiltext{3}{RIKEN (The Institute of Physical and Chemical Research),
                 2-1, Hirosawa, Wako, Saitama 351-0198, Japan}
\altaffiltext{4}{National Astronomical Observatory of Japan,
                 2-21-1 Osawa, Mitaka, Tokyo, 181-8588, Japan}


\begin{abstract}

A deep 75 ks {\it Chandra} ACIS--I data of NGC 2024 was analyzed, aiming 
at a search for diffuse X-ray emission in this one of the most nearby 
(415 pc) massive star-forming regions. After removing point sources, 
an extended emission was detected in the central circular region with a 
radius of 0.5 pc. 
It is spatially associated with the young massive stellar cluster. 
Its X-ray spectrum exhibits a very hard continuum ($kT>8$ keV)
and a sign 
of He-like Fe K$_\alpha$ line with the 0.5--7 keV absorption corrected 
luminosity of 2$\times10^{31}$ \ergs.
Undetected faint point sources, estimated from the luminosity function of
the detected sources, contribute less than 10\% to this emission.
Hence the emission is truly diffuse in nature. Because of the proximity of 
NGC 2024 and the long exposure, this discovery is one of the most strong 
supports for the existence of the diffuse X-ray emission in massive 
star-forming regions.

\end{abstract}




\section{Introduction}
\label{sec:intro}

Increasing evidences of diffuse X-ray emission 
has been found in massive star-forming regions (MSFRs) with 
the {\it Chandra X-ray Observatory}; 
the Rosette Nebula (at a distance of $D=1.4$ kpc; \citealt{Townsley2003}), 
M17 ($D=1.6$ kpc; \citealt{Townsley2003}), 
RCW 38 ($D=1.6$ kpc; \citealt{Wolk2002}), 
NGC 6334 ($D=1.7$ kpc; \citealt{Ezoe2006}),
RCW 49 ($D=$1.9-7.9 kpc; \citealt{Townsley2004}),
W51A ($D=$5.5-7.5 kpc; \citealt{Townsley2004}),
NGC 3603 ($D=7$ kpc; \citealt{Moffat2002}),
the Arches Cluster ($D=8.5$ kpc; \citealt{Yusef-Zadeh2002}), 
and the Quintuplet Cluster ($D=8.5$ kpc; ; \citealt{Wang2002}).
Townsley et al. (2003) explained the diffuse soft X-ray emission
found in M17 in the context of strong shocks by 
fast stellar winds from young massive stars
\citep{Dyson1972,Castor1975,Weaver1977}.
Recent results on NGC 6334 by \cite{Ezoe2006} indicate that
the spectra of the diffuse emission varies from place to place; 
those in tenuous molecular cloud regions are soft and thermal
with temperatures of several keV, while those in dense cores 
exhibit harder continua with a photon index of $\Gamma\sim$1.
They also have shown that these thermal and non-thermal spectra of 
the diffuse X-ray emission in MSFRs, found as a mixture in the 
NGC 6334 case, may be generally explained 
by the stellar-wind shock model.
In spite of these observational progresses, even with {\it Chandra},
there remains an uncertainty how much undetected faint point sources 
contribute to the emission, because these MSFRs are relatively 
distant ($>1.4$kpc).

In order to unambiguously examine the diffuse X-ray emission 
in MSFRs as possible as we can, we analyzed archival {\it Chandra} 
data of NGC 2024. 
This region, known as the Flame nebula, is one of the nearest MSFRs 
($D=415$ pc; \citealt{Anthony1982}), located in the Orion Nebula. 
It is an HII region considered to be illuminated by an O8V-B2V star 
IRS 2b \citep{Bik2003}. In the vicinity of IRS 2b, there are 
one early B star candidate associated
with an ultra compact HII region G206.543-16.347 and 
an infrared source IRS 2 \citep{Lenorzer2004}, 
and seven compact dust condensations named FIR 1-7, possibly 
massive protostars \citep{Mezger1988}. 
In addition, $\sim300$ low-mass 
($\simlt2$\msolar) young stars have been found 
by near infrared observations \citep{Lada1991}. 
The estimated age of NGC 2024 ranges from 0.3 \citep{Meyer1996} 
to several Myr \citep{Comeron1996}.
A previous analysis of the same {\it Chandra} data on point sources 
has been published by \cite{Skinner2003}. In this paper, we focus on 
a search for the diffuse X-ray emission in this representative MSFR.

\section{Observation}
\label{sec:obs}

{\it Chandra} observed NGC 2024 on 2001 August 8-9 
using the ACIS--I for 21.9 hr.
We started with the level 1 event files in the same way
as \cite{Skinner2003}. Utilized analysis software versions 
for the standard data reduction are different; we used the 
CIAO (Chandra Interactive Analysis 
of Observations) software version 2.3 and the calibration 
data base version 2.18.
These new softwares allowed us to correct the 
data for the effect of the charge transfer inefficiency, 
while it was not possible in \cite{Skinner2003}. 
No background flares were seen during the observation with the
average count rate of the six ACIS chips of 9.1 ct s$^{-1}$.
We then excluded $>1.2$ times of the average rate.
This procedure excluded 2\% of the exposure time.
After these data screenings, the nominal exposure 
has become 75.3 ksec. 

We searched the data for diffuse X-ray emission, 
following to the analysis method of \cite{Ezoe2006}. 
We first created adaptively smoothed 
X-ray images in two energy bands, as shown in figures 
\ref{fig:image} (a) and (b).
We see a sign of an extended emission associated with the massive
star IRS 2b and its vicinity. The extended emission in 0.5--2 keV is 
strong in the north-west direction of IRS 2b, while that in 2--7 keV 
is elongated in the north-west to the south-east direction.
We then used the {\tt wavedetect} program to identify sources 
using images in three energy bands (0.5--2, 2--8 and 0.5--8 keV) 
independently, in order not to miss very soft or strongly 
absorbed sources. The significance criterion and wavelet scales
were set at $1\times10^{-6}$ and 1-16 pixels in multiples of $\sqrt{2}$.
In the ACIS--I field of view, 301 sources were detected. Among 
them, 28 sources are newly identified ones, not listed in 
\cite{Skinner2003}, due to our three-band searching method. 

We excluded all the detected point sources by creating a point source 
mask using the ``Chandra Ray Tracer''$\footnote{http://cxc.harvard.edu/chart/threads/index.html}$.
For each source, we defined a radius to include $\sim98$ \% 
of photons at the Al K$\alpha$-line energy (1.497 keV).
Then we excluded all these regions by applying the mask to 
the raw ACIS--I image, and created images of the residual 
emission using the CIAO tools {\tt dmfilth} and {\tt csmooth}. 
Figures \ref{fig:image} (c) and (d) show thus obtained images 
of the extended emission.
In order to evaluate the significance of this emission,
we defined a circular region named C1 (the large circle in the panel d
of figure \ref{fig:image}) with a radius of 4$^\prime=$0.5 pc.
The total area of the C1 is 43/39 arcmin$^2$ or 0.62/0.52 pc$^2$ before/after
excluding the area around point sources. The 0.5--7 keV count rate
of the C1 is $0.142\pm0.001$ cts s$^{-1}$, while that of the same region
in blank-sky data is $0.084\pm0.001$ cts s$^{-1}$. 
Errors are 1$\sigma$ statistical ones.
Hence, its residual count rate is  $0.057\pm0.001$ cts s$^{-1}$ 
or 4270$\pm$110 cts. It is thus clear that a highly significant 
extended emission is present in the central region of NGC 2024.

\section{Extended X-ray Emission}
\label{sec:extended}

\begin{figure}
\centerline{
\includegraphics[width=0.35\textwidth,clip,angle=-90]{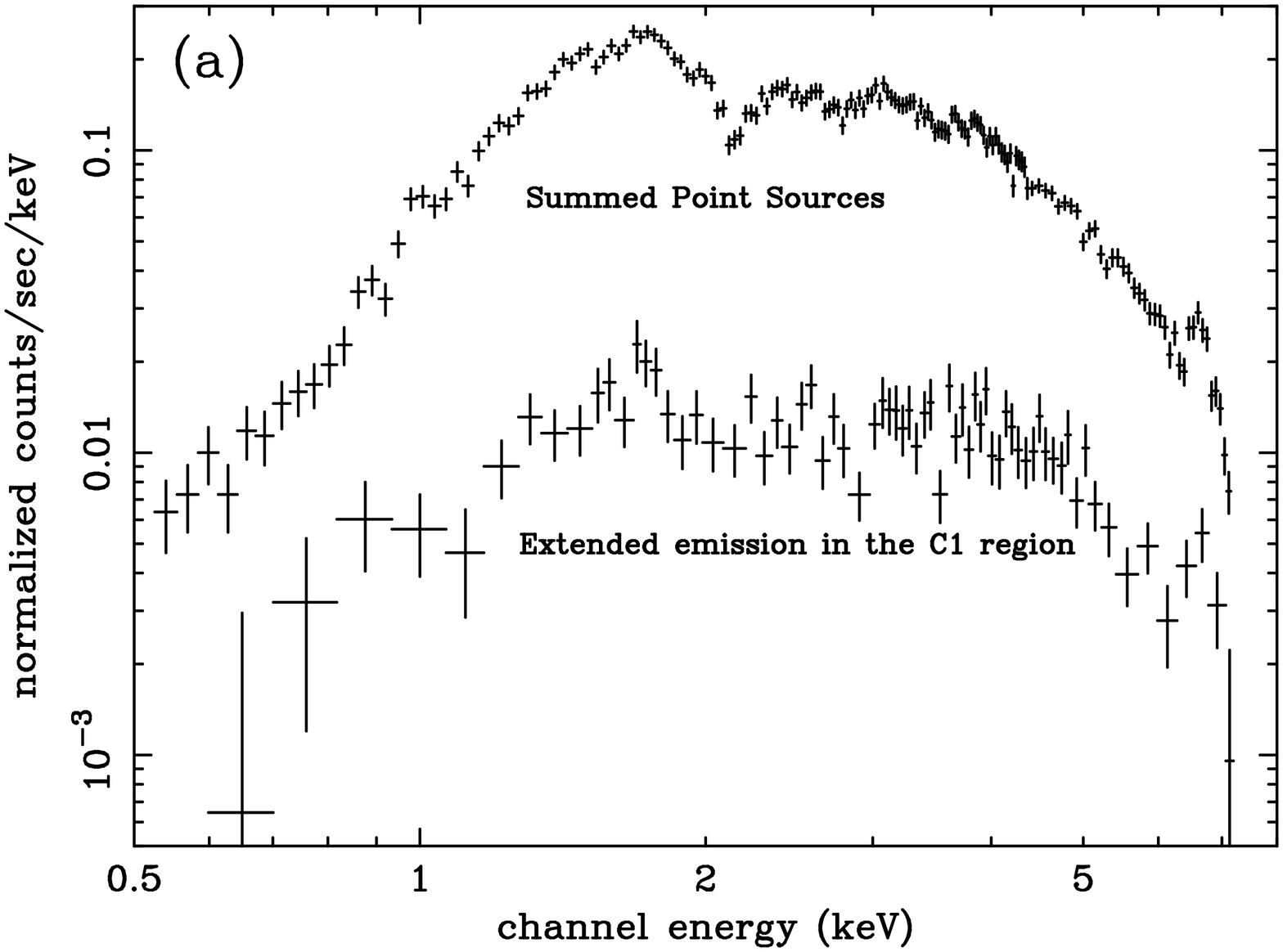}
\hspace*{0.02\textwidth}
\includegraphics[width=0.35\textwidth,clip,angle=-90]{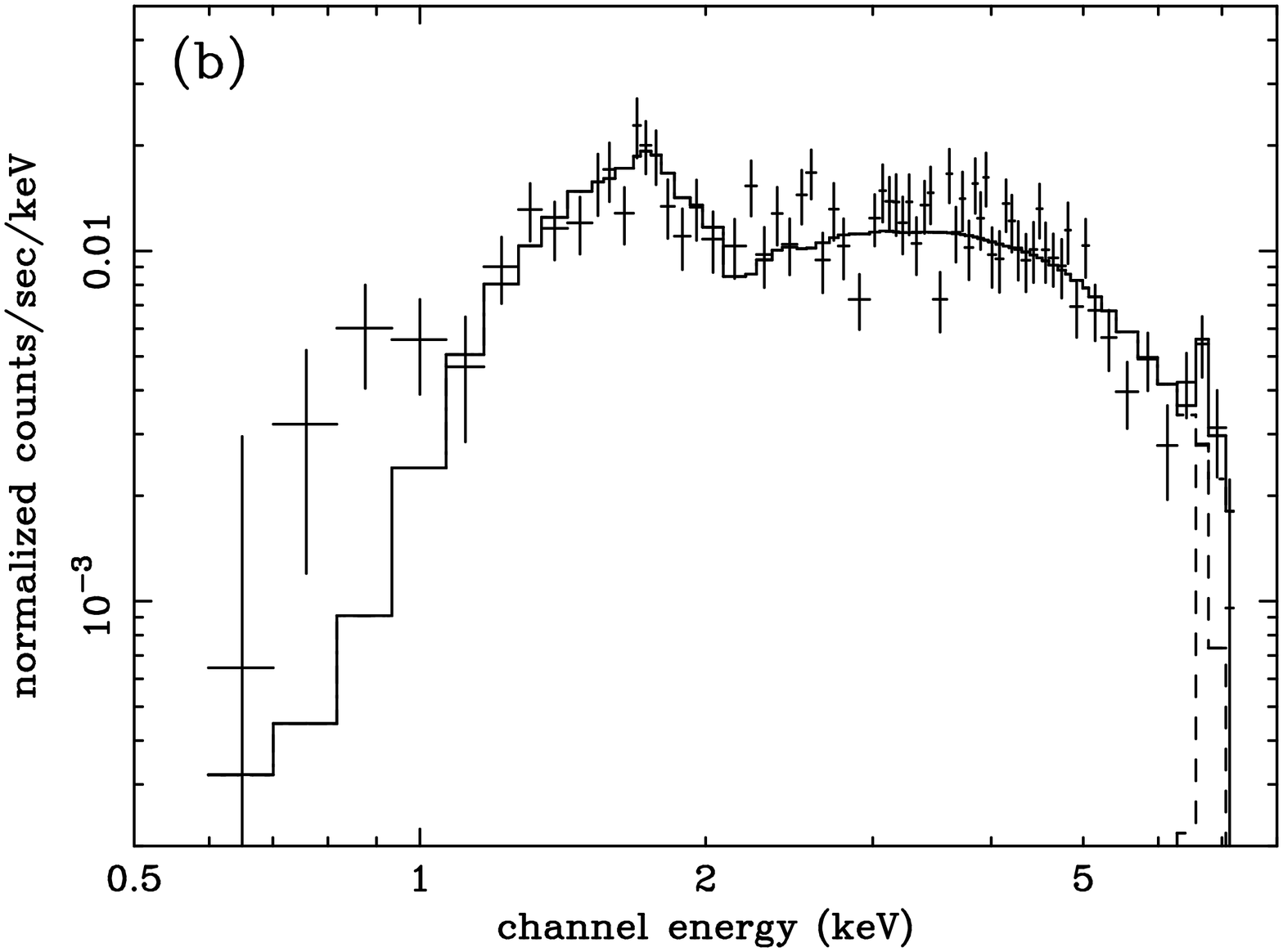}
}
\bigskip
\centerline{                                                         
\includegraphics[width=0.35\textwidth,clip,angle=-90]{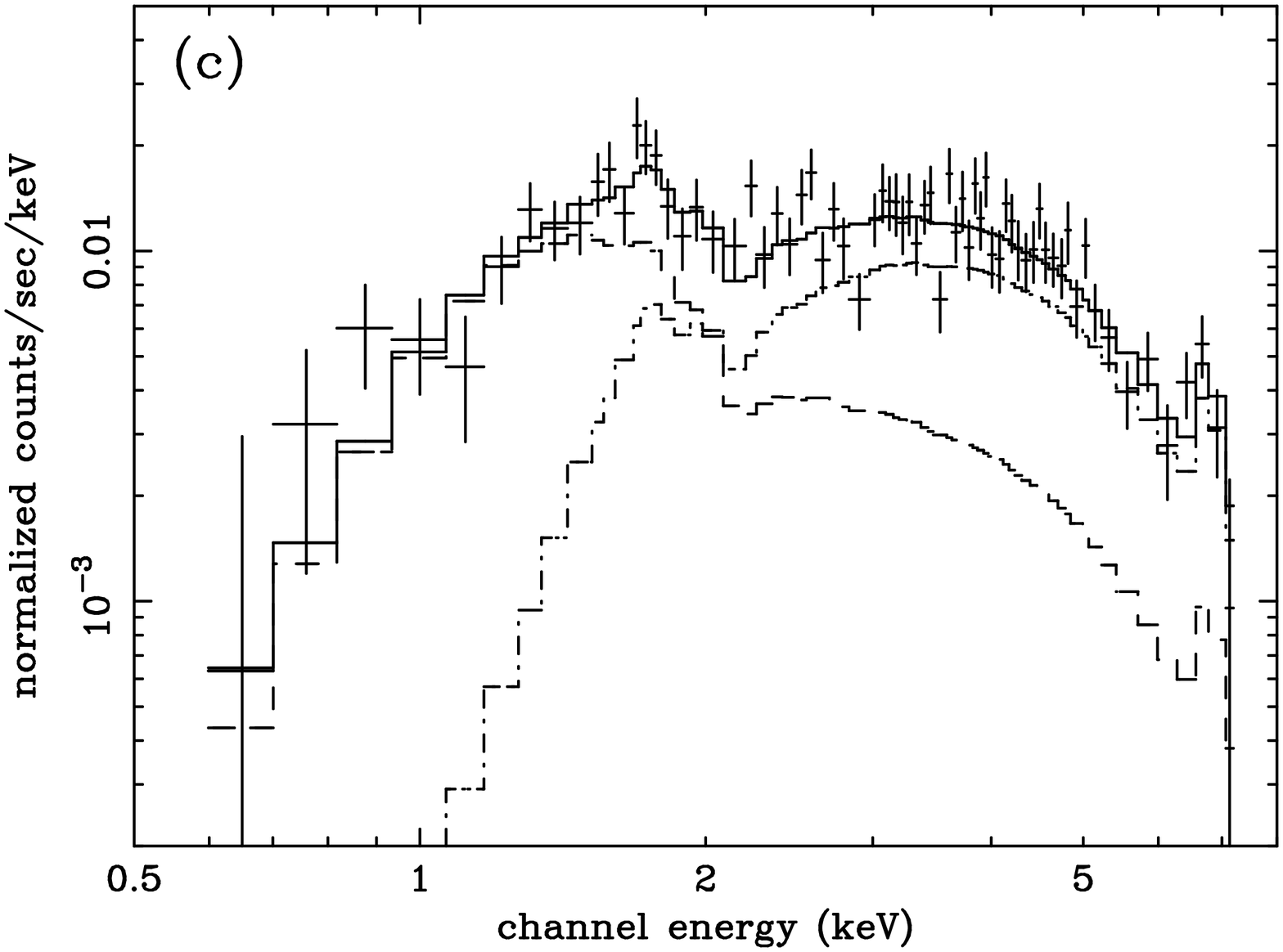}
\hspace*{0.02\textwidth}
\includegraphics[width=0.35\textwidth,clip,angle=-90]{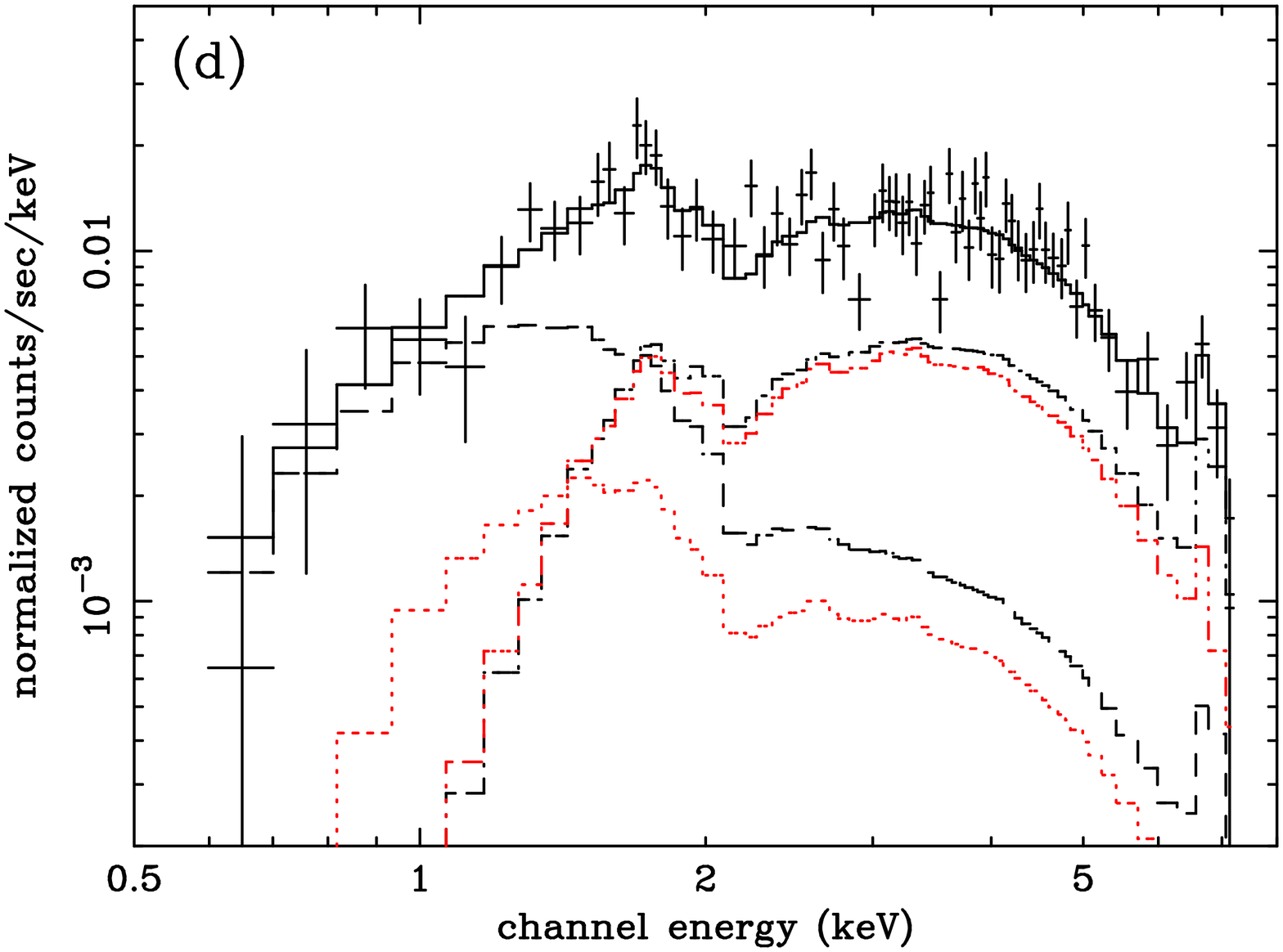}
}                                                                    
\caption{(a) ACIS spectrum of the C1 region compared with that 
         summed over 176 point sources.
         (b) The C1 spectrum fitted with a power-law plus a narrow
         Gaussian model.
         (c) The same as panel b, but for a single temperature plasma
         model suffering two absorption column densities (solid and dotted lines).
         (d) The same as panel c, but including the escaped photons from
         the summed point sources (red lines).
         }
\label{fig:spec}
\end{figure}

In the presence of the significant excess emission, 
we immediately considered a possibility of the diffuse emission. 
To know its basic properties, we compared its background subtracted 
spectrum in figure \ref{fig:spec} (a) with that summed over 176 
point sources detected within the C1. The weighted ARF (ancillary 
response function) and RMF (response matrix function) were calculated
using the CIAO programs {\tt mkwarf} and {\tt mkwrmf}, respectively.
The {\tt apply\_acisabs} script was utilized when creating ARF files,
to correct them for the decrease in the ACIS quantum efficiency.
The background spectra are extracted from the same regions in
the blank-sky data for individual regions. 
From figure \ref{fig:spec} (a), we can see important features of 
the extended emission. 
First, it is about one order of magnitude fainter than the 
summed point sources. Second, the extended emission show a harder 
continuum in the 1--7 keV range. Third, a hint of an emission line is 
seen in 6--7 keV.

To know basic parameters of the extended emission, we then conducted
spectral fitting. We employed a simple power-law model with an interstellar 
absorption and a narrow Gaussian. Here and hereafter all quoted errors
in the spectral fitting refer to 90\% confidence levels unless otherwise noted.
The result is shown in figure \ref{fig:spec} (b), where table \ref{tbl:fit} 
lists the obtained parameters.
The fit was not acceptable with $\chi^2/\nu\sim1.5$ because of the excess 
around 0.5--1 keV. The line center energy indicates a He-like Fe K$_\alpha$ 
line from thermal plasma. Then, we consider an alternative ``leaky absorber'' 
condition; a single thermal emission component reaches us via two (or more)
paths with different absorptions. 
This situation is possible in the MSFR like NGC 2024 where the density of 
the molecular cloud varies from place to place.
We hence fitted the spectrum by a sum of two thermal components with 
independent absorptions, but with their temperatures and abundances together.
The result is shown in figure \ref{fig:spec} (c) and table \ref{tbl:fit}.
As a thin thermal plasma model, we utilized the 
APEC (astrophysical plasma emission code)$\footnote{http://hea-www.harvard.edu/APEC/}$
model. Then, the fitting result is improved from $\chi^2/\nu\sim1.5$ to 1.3, 
which is significant with 99.7\% confidence according to an F-test.
The soft excess is represented by the mildly absorbed high temperature emission. 
The best fit temperature is high 11 keV, consistent with the small photon index 
of $\Gamma\sim0.9$ in the power-law model fit.
The 0.5--7 keV flux obtained from the leaky absorption model is 1.1$\times10^{-12}$ \flux.

\begin{figure}
\centerline{
\includegraphics[width=0.5\textwidth,clip,angle=-90]{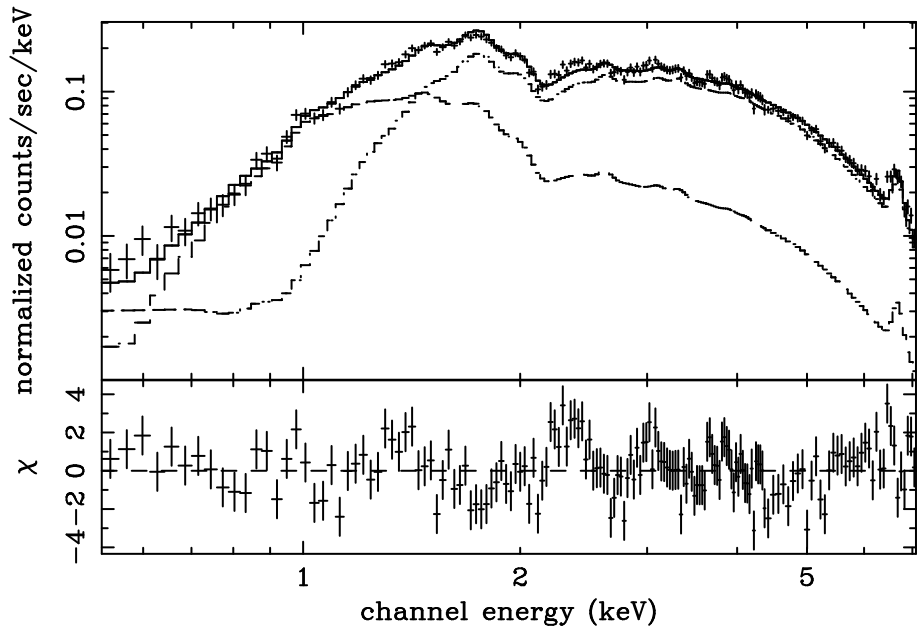}
}
\caption{The summed point-source spectrum in the C1 region, fitted
         with the leaky absorption model (solid and dotted lines).
         }
\label{fig:specsrc}
\end{figure}

For comparison, we quantified the summed point-source spectrum.
In the same manner of the extended emission, the source spectrum, 
the background, ARF, and RMF files were obtained. 
The fit for a single thermal emission model with one absorption
was not acceptable with $\chi^2/\nu\sim7$ because
this model cannot represent both soft excess below $\sim2$ keV, 
similar to that in the extended emission case, and also various 
emission lines. Hence, we used the leaky absorption
model with free Ne, Mg, Si, S, Ar, Ca, and Fe abundances, in order 
to represent the data better.
We obtained results as shown in table \ref{tbl:fitsrc} and 
figure \ref{fig:specsrc}. The best-fit temperature of 4.4 keV 
is consistent with typical values of young low-mass stars 
(e.g., \citealt{Imanishi2001a}) and significantly lower 
than that of the extended emission.

In spite of these spectral analysis, 
We here must evaluate an effect of photons escaping from 
the summed point sources because the point sources are
far brighter than the extended emission.
In the same way as \cite{Ezoe2006}, we 
took into account the summed point source spectrum 
by multiplying the best-fit 
model for the summed point sources by the energy-dependent 
escaping-fraction curve, estimated by the ``{\it Chandra} Ray Tracer'' (ChaRT). 
We have found that this contributes $\sim40$\% to the extended emission.
we remove their 98\% photons at 2 keV.
After correcting the escaping photon effect, the 0.5--7 keV flux of the 
extended emission becomes 6.3$\times10^{-13}$ \flux, yielding an 
absorption-uncorrected luminosity of 1.3$\times10^{31}$ \ergs or
the surface brightness of 2$\times10^{31}$ \ergs pc$^{-2}$ at 
an assumed distance of 415 pc.

We refitted the C1 spectrum including the escaping photon effect. 
The result is shown in figure \ref{fig:spec} (d) and table \ref{tbl:fit}. 
The fitting result was again acceptable. Also, the escaping photon effect 
has a relatively little effect on the fitting parameters except the 
normalizations. All the parameters are consistent with the previous ones 
before including this effect within 90\% errors. 
The absorption-corrected X-ray luminosity is 2$\times10^{31}$ \ergs.

In addition, we also conducted the same spectral analysis to 
a C2 region, the soft X-ray clump seen in figure \ref{fig:image} (c).
The obtained photon index and temperature are similar to 
those of the C1 region within errors, except a lower 
absorption column density ($0.1\times10^{22}$ cm$^2$ 
when fitted with the leaky absorption model including 
the escaping photon effect). 
No sign of emission lines was seen.
Hence, the spectral hardness of the extended emission is 
considered to be common within the whole region, and the 
offset peak in the soft band map is simply a consequence 
of a slight reduced absorption.

\section{Luminosity Function}
\label{sec:lf}

Based on the surface brightness of the extended emission, 
we estimated contribution from unresolved faint point sources
to this emission, in order to know whether it can be explained
by faint sources that are individually undetectable. 
We followed the same way of \cite{Ezoe2006} in which we utilized
the luminosity function of the detected point sources.
We below utilize the X-ray surface brightness of the extended 
emission after subtracting the escape photons from the point sources, 
obtained in the last section. 

\begin{figure}
\centerline{
\includegraphics[width=0.5\textwidth,clip,angle=-90]{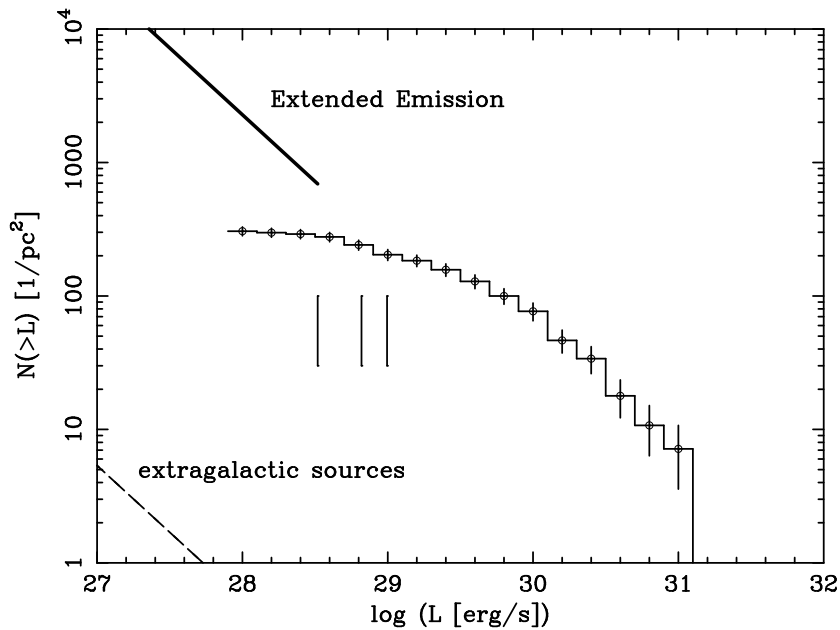}
}
\caption{ Luminosity function of the point sources within the C1.
          The vertical axis is the column density of the source number,
          while the horizontal axis absorption-uncorrected 0.5--7 keV
          luminosity. Errors are 1$\sigma$ Poisson. The thick line 
          indicates the number of putative point sources at a given
          flux that can account for the observed total surface brightness
          of the extended emission. Three short vertical lines are the
          typical X-ray fluxes of point sources with 10, 20 and 30 net 
          counts. The dashed line indicates the expected number of 
          extragalactic sources, estimated from \cite{Giacconi2001}.
         }
\label{fig:lf}
\end{figure}

Figure \ref{fig:lf} shows the luminosity function 
of the 176 point sources in the C1. We derived
the absorption-uncorrected 0.5--7 keV luminosity $L$.
The X-ray flux of each point source is obtained 
by spectral fitting for a bright source ($\geq30$ net counts
or counts after subtracting the background), while by using 
a count-to-flux conversion factor derived from the 
summed point sources 1.6$\times10^{-11}$ \flux (net counts/s)$^{-1}$
for a fainter source.

The source number density increases toward lower luminosities 
and saturates below 10 net counts, corresponding to the completeness 
limit of this observation. We here estimated the limit from source 
number histograms as a function of the logarithm of net counts, 
and regarded the maximum of the histogram as the completeness limit.
When utilizing the above conversion factor, 
10 net counts correspond to 4$\times$10$^{28}$ \ergs, in terms 
of the absorption-uncorrected 0.5--7 keV luminosity. 
This limit is one of the most lowest ones among those
in the past {\it Chandra} observations of MSFRs.

The solid line in figure \ref{fig:lf} indicates 
the necessary point source number, in order to 
explain the extended emission by point sources. 
The point number is clearly short, which supports 
that the extended emission is truly diffuse. 
Furthermore, even when we extrapolate the luminosity function 
toward a lower limit of 0 \ergs, using a linear function fitted in log-log space
from the completeness limit to $10^{30}$ \ergs, the estimated 
contribution of unresolved sources is at most $\sim10$\% of 
the extended emission. Based on these results, the extended 
emission of NGC 2024 can be considered as diffuse in nature.

\section{Discussion}
\label{sec:discuss}

We have found the diffuse X-ray emission in NGC 2024. 
Because of the proximity of NGC 2024 and the long exposure time, 
this discovery itself is one of the strongest supports for the existence 
of the diffuse emission in MSFRs, among the previous {\it Chandra} results.
Also, this result provides us with a new observational evidence that, 
even in a MSFR where only late O to early B stars exist, 
the diffuse X-ray emission can be observed if the sensitivity is enough high.
The spectral analysis suggests that the diffuse emission 
is dominated by the thermal emission. At the same time, since the continuum 
($kT>8$ keV) of the diffuse emission is harder than typical spectra of 
young stars, a part of the emission may come from non-thermal origin. 
In NGC 6334, the soft-thermal and hard possibly non-thermal regions can 
be spatially distinguished \citep{Ezoe2006}.
In NGC 2024, the soft and hard regions may be co-spatial and hence 
be observed as a mixture of both thermal and non-thermal emission.

\begin{figure}
\begin{center}
\includegraphics[width=0.7\textwidth,clip,angle=-90]{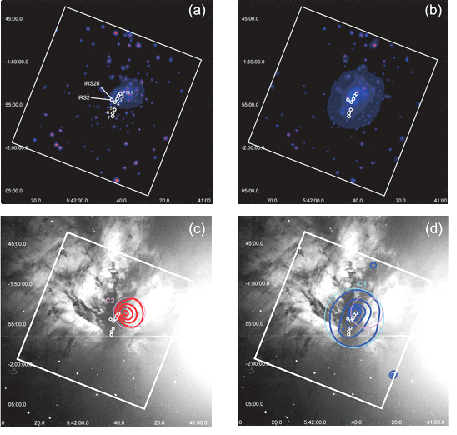}
\caption{Adaptively smoothed X-ray images of NGC 2024 taken 
         with the {\it Chandra} ACIS--I. 
         Panels (a) and (b) correspond to 0.5--2 keV and 
         2--7 keV band images before applying the point source mask, 
         respectively, while panels (c) and (d) are point-source
         excluded contours in 0.5--2 and 2--7 keV bands overlaid
         on the optical DSS image. 
         Co-ordinates are the J2000. All the images are 
         corrected for the exposure and vignetting, but the background 
         is not subtracted. 
	 Boxes show the ACIS--I fields of view, while large circles (C1 and C2)
         are regions utilized in our spectral analysis.
         Small circles are positions of IRS 2b, IRS2 and FIR 1-7.
         Because IRS 2 is located just 5$^{\prime\prime}$ south-east of IRS2b,
         these two circles are overlapped. 
         The color intensity is plotted logarithmically 
	 from 5$\times10^{-10}$ to 5$\times10^{-6}$
         counts s$^{-1}$ pixel$^{-1}$ cm$^{-2}$ in panel (a),
         while from 7$\times10^{-10}$ to 7$\times10^{-6}$ in panel (b).
	 Also the contours are plotted logarithmically from 
	 7.5$\times10^{-10}$ to 1.0$\times10^{-9}$
	 counts s$^{-1}$ pixel$^{-1}$ cm$^{-2}$ in panel (c),
         while from 1.3$\times10^{-9}$ to 4.0$\times10^{-9}$ in panel (d).
         }
\label{fig:image}
\end{center}
\end{figure}

We then discuss whether the thermal or non-thermal interpretation 
is feasible in terms of energetics. First, if we assume that the 
whole diffuse emission is thermal, the total plasma energy $U$ 
will be $\sim10^{47}\eta^{0.5}$\ergs where $\eta$ is a filling factor,
from the equation (3) in \cite{Ezoe2006}.  
Using the X-ray luminosity $L_{\rm X}$ and $U$, 
the plasma cooling time is estimated as $U/L_{\rm X}\sim10^8\eta^{0.5}$ 
yr, which is far longer than the age of NGC 2024, 
from 0.3 \citep{Meyer1996} to several Myr \citep{Comeron1996}. 
Although the sound crossing time in a 10 keV plasma across the region 
of 0.5 pc in size is $\sim10^3$ yr and hence short, the plasma with
the estimated pressure of $\sim5\times10^7 \eta^{-0.5}$ K cm$^{-3}$, 
calculated from the equation (5) in \cite{Ezoe2006}, may be confined 
by the dense HII region known to exist around the molecular cloud 
dark lane (figure \ref{fig:image} c)  \citep{Subrahmanyan1997}
and also the strong magnetic field 
within the molecular cloud \citep{Crutcher1999}. 
Then, by dividing the total energy by the assumed age of NGC 2024 of 
0.3 Myr, the average energy input is estimated as $10^{34}$ \ergs 

Second, we consider the non-thermal possibility.
The flat continuum ($\Gamma\sim1$ or $kT>8$ keV) of the emission strongly 
suggests the bremsstrahlung emission by 10 keV to several MeV electrons, 
rather than the synchrotron or inverse Compton emission \citep{Ezoe2006}. 
Since the Coulomb loss overwhelms the bremsstrahlung emission, if we assume 
that the diffuse emission is totally non-thermal, the necessary kinematic 
energy to be supplied is at least $\sim$10$^5$ times larger than the 
observed X-ray luminosity, $\sim$2$\times$10$^{36}$ \ergs. 

One of the most plausible energy sources for the diffuse emission is 
the shocks generated by fast stellar winds ($\sim2000$ km s$^{-1}$) 
from young massive stars. Its huge kinematic energy of the stellar 
winds can be easily converted via the strong shocks, among dense 
molecular clouds and HII regions, into the thermal (and non-thermal)
energy of particles in the surrounding gases.
This explanation has been proposed in M17 \citep{Townsley2003} 
and NGC 6334 \citep{Ezoe2006}.
In NGC 2024, there are at least one late O to early B star IRS 2b (O8V-B2V),
one early B star candidate IRS 2, 
and 7 possibly massive protostars. As shown in figure \ref{fig:image}, 
these massive stars are spatially associated with the diffuse emission. 
A typical kinematic luminosity of the stellar wind is 
$\sim10^{33-35}$ \ergs 
per one B2V-O8V star \citep{Howarth1989,Prinja1990,Panagia1973}. 
Therefore, the necessary energy supply of the thermal 
interpretation is explained if there is at least one 
massive star with a strong wind, comparable or stronger
than those of a typical B0.5V star.
The non-thermal interpretation is also possible if we consider 
all the 9 sources are massive stars and have strong winds. 
Wind-wind collisions may effectively increase the energy of a shock. 
Hence, the stellar wind scenario is possible from 
the viewpoint of the energetics.

%%Other observed parameters such as the plasma temperature and
%%the possible non-thermal component can be also explained by
%%the gas heating and particle acceleration in the stellar wind shocks.
%%Also, since the hot wind gas will expand since the energy supply of the
%%central OB star keeps constant, we can estimate the size 
%%of the expanding hot wind bubble as $\sim1$ pc, by assuming that 
%%the wind energy is equal to the displace energy of the cold gas 
%%surrounding the OB star. This estimated size is also consistent 
%%with the observed size. Hence, all these observed quantities 
%%support the stellar wind scenario.

YE is financially supported by 
the Japan Society for the Promotion of Science.

\clearpage

\clearpage

\clearpage

\clearpage

\clearpage

\begin{table}
\caption{Parameters of the spectral fitting to the extended emission 
         in the C1 region.}
\label{tbl:fit}
{\scriptsize
\begin{tabular}{lccccccccccccc} \hline\hline
       Model$^{\rm a}$  & $N_{\rm H1}$$^{\rm b}$   & $N_{\rm H2}$$^{\rm b}$ 
       & $\Gamma$ or $kT^{\rm c}$    & $Z^{\rm d}$ 
       & Norm.1$^{\rm e}$  & Norm.2$^{\rm e}$           
       & $E_{\rm C}^{\rm f}$         & Norm.$_{\rm line}^{\rm g}$ 
       & $F_{\rm X}^{\rm h}$         & $L_{\rm X}^{\rm i}$ 
       & $\chi^2$/d.o.f. \\\hline

       po+ga & 1.3$^{+0.5}_{-0.3}$ & $-$
       & 0.94$^{+0.32}_{-0.24}$ & $-$
       & 1.2$^{+0.6}_{-0.5}$ & $-$
       & 6.7$^{+0.1}_{-0.2}$    & 6.3$^{+4.6}_{-4.0}$ 
       & 1.1 & 2.9
       & 99.2/65 \\

       leaky abs.1  & 0.62$(<1.1)$ & 4.0$^{+1.4}_{-1.5}$
       & 11$^{+8.8}_{-3.5}$     & 1.3$^{+1.6}_{-0.8}$    
       & 2.2$^{+1.6}_{-1.7}$ & 9.2$^{+2.7}_{-1.6}$ 
       & $-$ & $-$
       & 1.1 & 4.0 
       & 85.9/64 \\

       leaky abs.2  & 0.21$(<1.1)$ & 3.3$^{+7.7}_{-1.0}$ 
       & 11($>$7.6)       & 2.5($>$1.0)
       & 0.73$^{+2.27}_{-0.43}$ & 4.4$^{+1.6}_{-1.4}$ 
       & $-$ & $-$
       & 0.63 & 2.2
       & 84.2/64 \\\hline

\end{tabular}
}
{\noindent
$^{\rm a}$ Fitting models. The po+ga, leaky abs.1 and leaky abs.2 indicate 
     the power-law plus a narrow Gaussian model, the leaky absorption model
     without and with the escaping photon effect from the point sources, respectively. \\
$^{\rm b}$ Interstellar absorption, with $N_{\rm H}$ being the hydrogen column density in 10$^{22}$ cm$^{-2}$. 
     In the case of the leaky absorption model, two column densities are given.\\
$^{\rm c}$ $\Gamma$ is the photon index in the power-law model, while $kT$ is a plasma temperature in keV.\\
$^{\rm d}$ Metal abundance in solar, used in the leaky absorption model. \\
$^{\rm e}$ Normalization is photon flux at 1 keV in 10$^{-4}$ photons cm$^{-2}$ s$^{-1}$ in the power-law model,
     and 10$^{-18}$/4$\pi D^2$ $EM$ in the leaky absorption model, where $D$ is a distance to NGC 2024 and 
     $EM$ is an emission measure in cm$^{-3}$.\\
$^{\rm f}$ A line center energy in keV.\\
$^{\rm g}$ Line intensity in 10$^{-6}$ photons cm$^{-2}$ s$^{-1}$. \\
$^{\rm h}$ The 0.5--7 keV flux in 10$^{-12}$ \flux. \\
$^{\rm i}$ The absorption-corrected 0.5--7 keV luminosity in 10$^{31}$ \ergs.
}
\end{table}

\clearpage

\begin{table}
\caption{Parameters of the spectral fitting to the summed point sources in the C1,
         derived by the leaky absorption model$^{\rm a}$.}
\label{tbl:fitsrc}
\begin{center}
\begin{tabular}{cccccccccccccc} \hline\hline
param & leaky abs. model \\\hline
$N_{\rm H1}$ & 0.55$\pm0.09$  \\
$N_{\rm H2}$ & 2.7$^{+0.3}_{-0.2}$  \\
$kT$         & 4.4$^{+0.3}_{-0.4}$  \\
$Z_{\rm Ne}$$^{\rm b}$ & 2.9$(>1.2)$  \\
$Z_{\rm Mg}$$^{\rm b}$ & 1.1$\pm0.8$  \\
$Z_{\rm Si}$$^{\rm b}$ & 0.06$(<0.32)$  \\
$Z_{\rm S} $$^{\rm b}$  & 0.93$^{+0.37}_{-0.36}$  \\
$Z_{\rm Ar}$$^{\rm b}$ & 1.2$^{+0.7}_{-0.5}$  \\
$Z_{\rm Ca}$$^{\rm b}$ & 0.59$(<1.2)$  \\
$Z_{\rm Fe}$$^{\rm b}$ & 0.23$^{+0.04}_{-0.03}$  \\
Norm.1       & 17$^{+5}_{-4}$  \\
Norm.2       & 130$\pm10$  \\
$F_{\rm X}$  & 9.5  \\ 
$L_{\rm X}$  & 43  \\
$\chi^2$/d.o.f. & 292.1/157\\\hline
\end{tabular}
\end{center}
{\noindent
$^{\rm a}$ Notations and symbols are the same as the leaky absorption model in table \ref{tbl:fit} 
           except abundances.\\
$^{\rm b}$ Metal abundances of individual elements in solar, while those of other elements
           are fixed at 1 solar.\\
}
\end{table}

\end{document}